# Light-Induced Static Magnetization: Nonlinear Edelstein Effect

Haowei Xu [1], Jian Zhou [1], Hua Wang [1], and Ju Li [1,2] *

[1] Department of Nuclear Science and Engineering, Massachusetts Institute of Technology, Cambridge, Massachusetts 02139, USA

[2] Department of Materials Science and Engineering, Massachusetts Institute of Technology, Cambridge, Massachusetts 02139, USA

---

* H.X. haoweixu@mit.edu  J.Z. jianzhou.pku04@gmail.com  H.W. daodao@mit.edu  J.L. liju@mit.edu

Correspondence to: J.L. liju@mit.edu  Phone: 617-253-0166

**Abstract**

Light can interact with magnetism in materials. Motivated by the Edelstein effect, whereby a static electric field can generate magnetization in metals, in this work we theoretically and computationally demonstrate that static magnetization can also be generated through light in semiconductors. Such an effect is essentially a second-order nonlinear response and can be considered as a generalization of the Edelstein effect. This nonlinear Edelstein effect (NLEE) applies to semiconductors under both linearly and circularly polarized light, and there are no constraints from either spatial inversion or time-reversal symmetry. With *ab initio* calculations, we reveal several prominent features of NLEE. We find that the light-induced orbital magnetizations can be significantly greater than the spin magnetizations, in contrast to standard intrinsic magnetism where the orbital magnetic moment is strongly quenched under crystal field. We show that in multi-layer (multi-sublattice) materials, different ferromagnetic and ferrimagnetic structures can be realized under photon pumping, depending on the inter-layer (inter-sublattice) symmetry. It is also possible to switch the magnetic ordering in anti-ferromagnetic materials. The relationship between NLEE and other magneto-optic effects, including the inverse Faraday effect and inverse Cotton-Mouton effect, is also discussed.

**Introduction**

The generation and manipulation of magnetization is the basis of magnetic information storage and spintronics [1,2]. Conventionally, one uses an external magnetic field to read and write magnetism. However, since the magnetic coupling is weak, a strong magnetic field is usually necessary. For example, recent experimental works show that ~ 1 T magnetic field is required to flip the CrI$_3$ bilayer from anti-ferromagnetic into ferromagnetic ordering [3]. Furthermore, in contrast to electric fields or light beams, it is hard to spatially confine and focus the magnetic fields. Modern spintronics requires fast and precise control of magnetization, and one may have to resort to electrical or optical approaches. Physically, it has been demonstrated that an optical or electric field with moderate strength could induce sufficient effective magnetic field and control the magnetism in both bulk materials and thin films efficiently [4–7].

Maxwell's equations in vacuum couple electric field with magnetic field in a standard way, but it would be desirable to further couple magnetization with electric fields by interacting with a materials medium. Indeed, a static electric field can generate and manipulate magnetization. A typical example is the linear Edelstein effect (LEE). LEE is essentially a conversion between electric field and magnetization: in a non-centrosymmetric metal, a static magnetization $M$ can be generated when a static electric field $E$ is applied, and one has $M^i = \zeta_a^i E^a$, where $\zeta_a^i$ is the response function, while $i$ and $a$ indicate the polarization of the magnetization and the electric field, respectively. LEE was first theoretically proposed [8] and then experimentally realized [9,10]. Recently, there are growing interests in LEE [5,11–13], and it was suggested that LEE can potentially switch the magnetic orderings of magnetic materials [5,13]. However, LEE only exists in metallic systems, and the electric field would also generate a charge current (Ohm current). As a result, sometimes the Edelstein effect is also described as the conversion between charge current and magnetization. As we will elaborate later, the Edelstein effect and Ohmic current have a similar physical origin and can be regarded as cousin processes. Another effect that can convert electric field to magnetization is the magnetoelectric effect [14], which can also be described by $M^i = \alpha_a^i E^a$. Magnetoelectric effect works in insulating materials, but it requires the breaking of both time-reversal and spatial inversion symmetry. In the following we would focus more on non-magnetic materials, i.e. with time-reversal symmetry, at the ground state.

Besides a static electric field, we will show how light can be used to manipulate magnetization. As electromagnetic waves, light has alternating electric and magnetic field components, both of which can interact with magnetization. The first observation of the interaction between light and magnetism (magneto-optical effect) dated back to the 1840s, when Faraday experimentally discovered that the polarization plane of a linearly polarized light would be rotated when light propagates in magnetic materials. This phenomenon, dubbed the Faraday effect (and a related magneto-optic Kerr effect in reflection mode), vividly demonstrates that magnetism has an influence on light, and then finds wide applications, such as the measurement of magnetism. In recent years, as lasers with high intensity become available, the inverse effects start to attract great attention, which may have applications in ultrafast spintronics and data storage, etc. The magnetic field component can directly interact with magnetic moments through Zeeman coupling [15]. But the Zeeman coupling is usually weak, so of greater interest is the coupling between the electric field component with magnetization. For example, Ref. [16] made use of the inverse Faraday effect (IFE) and showed that a circularly polarized light (CPL) can generate an effective magnetic field, which can non-thermally manipulate the magnetization of magnetic materials. Later, it was shown [17] that even a linearly polarized light (LPL) can cause coherent spin excitations through the inverse Cotton-Mouton effect (ICME). The optical controls over magnetism are extensively studied these days [4,7]. In this work, we develop a computational approach to this problem.

The LEE is a first-order response to the electric field ($\delta M \propto E$). In this work, we generalize LEE into the second-order nonlinear Edelstein effect (NLEE). For NLEE, a static magnetization $\delta M \propto E(\omega)E(-\omega)$ is generated under light. $E(\omega)$ is the Fourier component of the oscillating electric field of the light at angular frequency $\omega$. Just as LEE and Ohmic currents are cousins, NLEE is the cousin process of the bulk photovoltaic effect [18–20], via which a static charge current is generated under light. We find that the NLEE can induce a larger effective magnetic field than that of LEE at moderate electric field strength ($E \gtrsim$ 10 MV/m). The strength of the magnetization generated by NLEE depends linearly on the light intensity, and can be detected by quantum sensors [21] such as SQUID, NV centers, etc., even if the light intensity is mild so that the magnetization generated is too small to be detected by conventional approaches such as magneto-optical Kerr rotation. From symmetry considerations, NLEE is not constrained by either spatial inversion $\mathcal{P}$ or time-reversal $\mathcal{T}$ symmetry, whereas LEE vanishes in $\mathcal{P}$-conserved systems. In addition, since light can induce electron inter-band transitions, NLEE can exist in semiconductors and insulators, whereas LEE only exists in metallic systems. Hence, NLEE can be active in many more materials compared with LEE. Notably, magnetization can be generated under LPL in nonmagnetic materials. This is somewhat counter intuitive, as magnetization requires $\mathcal{T}$-breaking while LPL cannot break $\mathcal{T}$. We attribute this effect to the breaking of $\mathcal{T}$ by energy dissipation in the photocurrent generated by the above-bandgap photon absorption. Furthermore, as an optical effect, NLEE enjoys many salient merits of optical approaches, as it can be non-contact, non-invasive, and ultrafast. These factors render NLEE a potentially effective method for generating and manipulating magnetic structures, including ferro-, ferri- and anti-ferromagnetism.

In the following, we first introduce the physical mechanism and theory of NLEE. Then to illustrate some prominent properties of NLEE, we perform *ab initio* calculations in different material systems, including non-magnetic transition metal dichalcogenides (TMD), e.g., $MoTe_2$ and antiferromagnetic $CrI_3$ bilayers. We incorporate orbital magnetic moments, as well as spin magnetic moments. Remarkably, we find that the orbital contribution can be stronger than the spin contribution, especially in conventionally nonmagnetic systems. This is opposite to the behavior of the spontaneous magnetization of magnetic materials, where the spin contribution usually dominates. In bilayer $MoTe_2$, NLEE is sensitive to the stacking pattern of the two $MoTe_2$ layers, and various opto-magnetic orderings, including anti-ferromagnetic (AFM) and ferromagnetic (FM), are achievable and controllable. Finally, we discuss the possibility of switching the AFM ordering in the $CrI_3$ bilayer with NLEE, making use of the spatially varying magnetization. The relationship between NLEE and IFE and ICME is also addressed. Specifically, IFE and ICME are incorporated by NLEE, and NLEE also points out the possibility to generate magnetization in non-magnetic materials under LPL, which is not captured by IFE or ICME, and to the best of our knowledge, has not be proposed before.

## Results

**Mechanisms and Theory.** The electron magnetic moment $m$ has both the spin ($S$) and orbital ($L$) angular momentum contributions, and one has $m = \mu_B(2S + L)/\hbar$, where $\mu_B$ is the Bohr magneton and the factor of 2 for $S$ is the $g$-factor of the electron spin. The total magnetization of an electron ensemble is the total magnetic moments of all electrons. For example, in nonmagnetic materials, the magnetic moments of all electrons sum up to zero, and the equilibrium magnetization $M_0$ is thus zero. However, when the electrons are driven out of equilibrium, the electronic distribution function would be changed, and a net magnetization $\delta M$ may arise. A simple example is, in a system where the magnetic moments of all electrons cancel out, if somehow one electron flips its magnetic moment, then the system would acquire a net magnetization. One can see that for the total magnetization to be nonzero, each electron should have a specific magnetic moment (i.e., spin texture and orbital texture). If the magnetic moments of the electrons are random, then the total magnetization would always be zero, no matter what the distribution function looks like. For the spin part, the spin texture could be created by, e.g., spin-orbit coupling (SOC) or magnetic ordering. For the orbital part, the orbital texture is ubiquitous in multi-orbital systems, and a nonzero orbital magnetization generally exists.

Various mechanisms can drive the electrons out of equilibrium, thus change the magnetic state. For the LEE, it is the electric field $E$ that drives the electrons out of equilibrium. Indeed, electrons would move under the electric field, leading to a change in the distribution function. This effect is schematically illustrated in Figure 1a: Region X (Y) of the Brillouin zone will have fewer (more) electrons under $E$, which tilts the Fermi surface. If electrons in X and Y have different spin/orbital polarizations $m^X \neq m^Y$, then a net magnetization change $\delta M \propto m^Y - m^X$ would arise. One can see that the LEE is an *intraband* process and only electrons near the Fermi level contribute – thus the LEE applies only to metals. LEE has a similar physical origin to the Ohmic current. At equilibrium, the velocities of all electrons sum up to zero, thus the net charge current is zero. But the electrons in regions X and Y have different velocity $v^X \neq v^Y$. As a result, when the electric field is applied, the velocities of all electrons do not sum up to zero anymore, and a net charge current $j \propto v^Y - v^X$ would be generated.

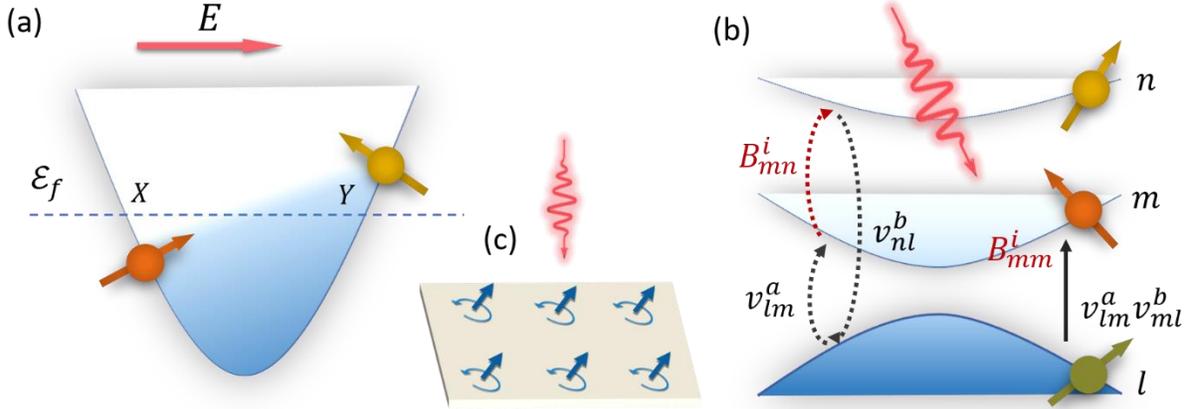

**Figure 1.** Simplified physical pictures of (a) linear Edelstein effect and (b) nonlinear Edelstein effect. The blue filling indicates electron occupation. For linear Edelstein effect, the static electric field modifies electron distribution in a single band (intraband process). For nonlinear Edelstein effect, light excites interband transitions between different bands (labeled with $l, m, n$). The three-band process in Eq. (2) is illustrated by the dashed arrows on the left side of (b), while the two-band process in Eq. (4) is illustrated by the solid arrow on the right side. (c) Local vortices in the photocurrent (blue curved arrows) and the associated orbital magnetization (blue straight arrows) under light (red wavy arrow) illumination.

We now extend from static electric field to optical alternating electric field and generalize the first-order LEE to the second-order NLEE. The magnetization induced by an alternating electric field can be expressed as,

$$\delta M^{i,\beta} = \sum_{\Omega=\pm\omega} \chi_{ab}^{i,\beta}(\Omega) E^a(\Omega) E^b(-\Omega) \qquad (1)$$

Here $a, b$, and $i$ indicate the directional component of the electric field and the magnetization, respectively. $E(\omega)$ is the Fourier component of the electric field at angular frequency $\omega$. $\chi_{ab}^{i,\beta}(\omega)$ is the nonlinear response tensor. Superscript $\beta$ indicates either spin ($\beta = S$) or orbital ($\beta = L$) degree of freedom, and a total ($\beta = T$) magnetic moment is $\delta M^{i,T} = \delta M^{i,S} + \delta M^{i,L}$. Eq. (1) suggests that when the $\omega$ and $-\omega$ frequency components of the light's electric field are combined, a static magnetization is generated. This is similar to the difference frequency generation and bulk photovoltaic effect. For the difference frequency generation, two photons with frequencies $\omega_1$ and $\omega_2$ are combined, and a third photon with frequency $\omega_1 - \omega_2$ is generated. For the bulk photovoltaic effect, the $\omega$ and $-\omega$ frequency components of the electric field are combined, and a displacement of electrons in real space (charge current) is generated. For NLEE, instead of a third photon or an electron displacement, a static magnetic moment is generated. The NLEE magnetization is characterized by the response function $\chi_{ab}^{i,\beta}$, which will be the focus in the following. The formula of $\chi_{ab}^{i,\beta}$ can be obtained from quadratic response theory [18,22–25]. Within the independent particle approximation, it can be expressed as

$$\chi_{ab}^{i,\beta}(\omega) = -\frac{\mu_B e^2 V_{u.c.}}{\hbar^2 \omega^2} \int \frac{d\mathbf{k}}{(2\pi)^3} \sum_{mnl} \frac{f_{lm} v_{lm}^a}{\omega_{ml} - \omega + i/\tau} \left( \frac{\beta_{mn}^i v_{nl}^b}{\omega_{mn} + i/\tau} - \frac{v_{mn}^b \beta_{nl}^i}{\omega_{nl} + i/\tau} \right) \quad (2)$$

We have omitted the $\mathbf{k}$-dependence of the quantities in the integrand. $\mu_B$, $e$ and $\hbar$ are the Bohr magneton, electron charge, and reduced Planck constant, respectively. Here we multiply unit cell volume ($V_{u.c.}$), so that $\chi_{ab}^{i,\beta}$ corresponds to the magnetization in a unit cell, rather than a magnetization density. $f_{lm} \equiv f_l - f_m$ and $\hbar \omega_{lm} \equiv \hbar(\omega_l - \omega_m)$ are the difference between equilibrium occupation number and band energy between bands $|l\rangle$ and $|m\rangle$, respectively. $\mathbf{v}_{nl} = \langle n|\hat{\mathbf{v}}|l\rangle$ is the velocity matrix. For the spin and orbital contributions, one can set $\boldsymbol{\beta}_{mn} = 2\mathbf{S}_{mn} = 2\langle m|\hat{\mathbf{S}}|n\rangle$ and $\boldsymbol{\beta}_{mn} = \mathbf{L}_{mn} = \langle m|\hat{\mathbf{L}}|n\rangle$, where $\hat{\mathbf{S}}$ and $\hat{\mathbf{L}}$ are spin and orbital angular momentum operators, respectively. The carrier lifetime $\tau$ is assumed to be a constant for all electronic states and is set as 0.2 ps in the following. The dependence of $\chi_{ab}^{i,\beta}$ on $\tau$ can be found in [25]. We define symmetric real and asymmetric imaginary parts of $\chi_{ab}^{i,\beta}$ as

$$\eta_{ab}^{i,\beta} \equiv \frac{1}{2} \text{Re} \left\{ \chi_{ab}^{i,\beta} + \chi_{ba}^{i,\beta} \right\}$$
$$\xi_{ab}^{i,\beta} \equiv \frac{1}{2} \text{Im} \left\{ \chi_{ab}^{i,\beta} - \chi_{ba}^{i,\beta} \right\} \quad (3)$$

Note that $\eta_{ab}^{i,\beta}$ and $\xi_{ab}^{i,\beta}$ correspond to the response function under LPL and CPL, respectively.

From Eq. (2) one can see that light can excite (virtual) inter-band transitions of electrons, as illustrated in Figure 1b. The virtual transition between band $m$ and $l$ is mediated by $n$. When at equilibrium, the electron tends to occupy states with the lowest energy, so they should reside on the $l$-th band. Under light illumination, the electron can (virtually) transit to the $m$-th band, which has higher energy. If on a $\mathbf{k}$-point, different bands have different spin/orbital polarization, then a net magnetization change $\delta M^i$ can be established. Similarly, (virtual) electrons interband transition is also the foundation of difference frequency generation and bulk photovoltaic effect. One could use symmetry analysis to examine the response explicitly. Under spatial inversion $\mathcal{P}$, axial vectors $M$, $S$ and $L$ are even, while polar vectors $E$ and $v$ are odd. From both Eqs. (1) and (2), one can deduce that the NLEE does not require $\mathcal{P}$ breaking. This is in contrast to LEE, difference frequency generation, or the bulk photovoltaic effect, which vanish in $\mathcal{P}$-conserved systems. It is more intriguing to study the NLEE under time reversal operation $\mathcal{T}$. Under $\mathcal{T}$ operation, $M, S, L$ and $v$ are odd, while $E$ is even. From Eq. (1) one may deduce that the NLEE should vanish in a $\mathcal{T}$-conserved system under LPL: $\mathcal{T}M = -M$ and $\mathcal{T}E = E$, leading to $\mathcal{T}\chi = -\chi$, thus $\chi$ needs to be zero to preserve time-reversal symmetry. However, Eq. (2) would yield a contrary conclusion: $\mathcal{T}$ does not enforce a zero $\chi_{ab}^{i,\beta}$ (see detailed analysis in [25]). Intuitively, LPL induces photocurrent with vorticity as it flows pass atoms with chiral neighboring surroundings (crystal field), like eddy when water flows past rocks in a stream. The vortex currents generally lead to a net magnetization. In $\mathcal{P}$-broken systems, the vortex currents do not exactly cancel, and lead to a net charge current in the bulk, which is the bulk photovoltaic effect. In $\mathcal{P}$-conserved systems, the net charge current should vanish in the bulk, but there is still a net current on the surfaces, where $\mathcal{P}$ is naturally broken. The contradiction between Eqs. (1) and (2) can be resolved if one considers the dissipation. Light with above-bandgap frequencies can be absorbed by electron interband transitions and then be dissipated as heat. Such dissipation breaks $\mathcal{T}$ of the light-matter system, according to the second law of thermodynamics. Similar reasonings apply to the Ohmic current. Under $\mathcal{T}$, the charge current is odd ($\mathcal{T}j = -j$), while the electric field is even. But the Ohmic current does exist. This is because the Joule heat breaks $\mathcal{T}$, even if the material possesses $\mathcal{T}$ in equilibrium. Actually, the LEE does not require explicit $\mathcal{T}$-breaking either, and the conversion between charge current and magnetization can happen in non-magnetic materials. This is also because the Joule heat associated with the charge current breaks $\mathcal{T}$.

Here we would like to discuss further the role of the carrier lifetime $\tau$. In the hypothetical "clean limit" where no dissipation exists ($\tau \to \infty$), the NLEE tensor should be zero under LPL in a $\mathcal{T}$-symmetric system, based on the symmetry analysis on Eq. (1). Technically, this is also manifested in Eq. (2), and the reason is as follows. Under time-reversal $\mathcal{T}$ operation, one has $\mathcal{T}v_{mn}(\mathbf{k}) = -v_{mn}^*(-\mathbf{k})$ and $\mathcal{T}\beta_{mn}(\mathbf{k}) = -\beta_{mn}^*(-\mathbf{k})$, where $*$ indicates the complex conjugate. Thus, the numerator, which is $N_{mnl} = v_{mn}v_{nl}\beta_{lm}$, would behave as $\mathcal{T}N_{mnl}(\mathbf{k}) = -N_{mnl}^*(-\mathbf{k})$. After the summation over $\pm\mathbf{k}$, the numerator would be purely imaginary. If $\tau = \infty$ and $\frac{i}{\tau} = 0$, then the denominator would be purely real. Therefore, the whole formula is purely imaginary, and cannot contribute to a static magnetization, which should be a real number. From this point of view, Eqs. (1) and (2) are mathematically equivalent even in the "clean limit". But in practice, $\tau$ cannot and should not go to infinity. Mathematically, Eq. (2) may experience divergence problem if one sets $\frac{\hbar}{\tau} = 0$, so a finite $\hbar/\tau$ is necessary. Such phenomenon is common in e.g., quantum field theory, where one adds a small but finite imaginary term in the propagator to avoid the divergence at the singular point. Physically, in interacting systems $\hbar/\tau$ has a physical meaning of (effective) self-energy, and $\tau$ is the

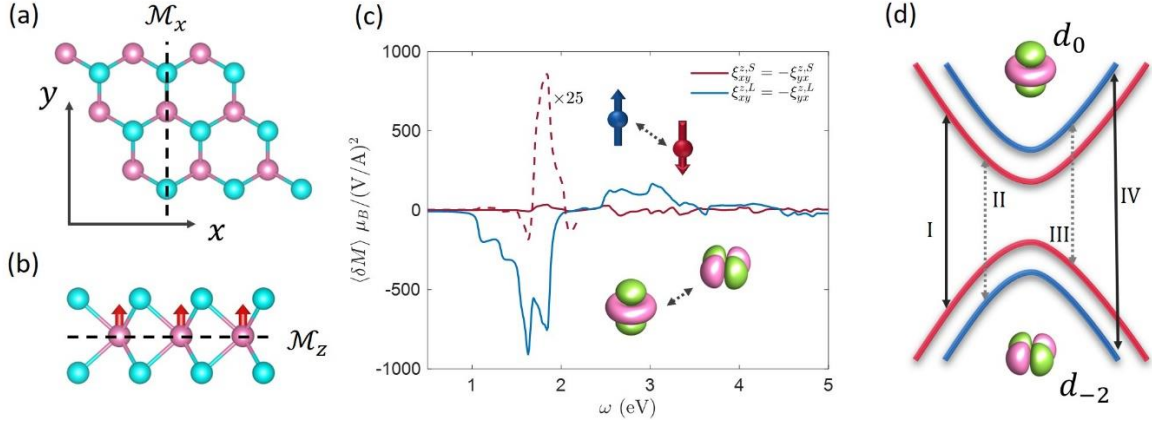

**Figure 2** (a, b) Top and side view of monolayer MoTe$_2$. The mirror symmetries are indicated by dashed line. The red arrows in (b) denotes the magnetization under CPL. (c) Orbital (blue) and spin (red) contributions to the total nonlinear Edelstein effect under circularly polarized light. The dash red curve is $\xi_{xy}^{z,S}$ amplified by 25 times for $\omega < 2.2\ eV$. The magnetization is shown for a primitive cell. (d) Schematic spin and orbital projected band structure of MoTe$_2$ near K point. Blue (red) indicates spin up (down) states, while the valance and conduction bands have major contribution from $d_{-2}$ and $d_0$ orbitals, respectively. I-IV denote four possible interband transitions.

lifetime. The electrons are never free particles in a solid-state system, and their line width ($\hbar/\tau$) is always finite. Even if the sample is a perfect crystal, there are still electron-electron interactions, electron-phonon interactions, etc. Besides, the application of Eq. (2) and Eq. (4) in the following requires extra care. One should treat Eqs. (2) and (4) as low-order perturbation theory and should not apply them when the system is strongly out-of-equilibrium.

In the following, we perform *ab initio* calculations to illustrate NLEE in various 2D materials that have a large surface area-to-volume ratio and are easily accessible with optical pumping. We first use monolayer TMDs as an example to show that the orbital contribution to the magnetic moment can be significantly greater than the spin contribution in intrinsic nonmagnetic systems. Then we use bilayer TMDs to show that different magnetic orderings can be obtained under LPL, depending on the interlayer stacking symmetry. Finally, we take 2D AFM material CrI$_3$ to discuss the possible AFM order manipulation under NLEE. We use 2D materials because they are simpler than 3D materials and various features of NLEE can be better illustrated. The theory of NLEE applies to 3D materials as well, and can generate a larger total magnetic moment in e.g., thin films and conventional 3D bulks.

**Monolayer MoTe$_2$: Spin and Orbital Contributions**. For the spontaneous magnetization in magnetic materials, the contribution from the orbital angular momentum $L$ is usually weaker than that from the spin angular momentum $S$. Typically the orbital contribution is less than 10% of the total magnetization $m = \mu_B(2S + L)/\hbar$ [26]. This is due to the orbital quenching by a strong crystal field. On the contrary, for the non-equilibrium magnetization, we will show that the orbital angular momentum could contribute more significantly than the spin angular momentum, due to the chirality of the same strong crystal field.

As an example, we use TMDs (MoTe$_2$) in 2H phase, which exhibit many peculiar properties and are widely studied in recent years. Monolayer 2H TMDs possesses mirror symmetries $\mathcal{M}_x$ and $\mathcal{M}_z$, as indicated by the dashed lines in Figures 2a-b. Notably, $\mathcal{M}_z$ enforces Zeeman type (out-of-plane) spin/orbital texture. Here we need to examine the constraints on NLEE from mirror symmetries. The polar vector $\boldsymbol{E}$ satisfies $\mathcal{M}_j E^a = (-1)^{\delta_{ja}} E^a$, where $\delta_{ja}$ is the Kronecker delta. That is, the $j$-th component of $\boldsymbol{E}$ is flipped under $\mathcal{M}_j$. On the other hand, the spin or orbital angular momentum $\boldsymbol{\beta}$ is an axial vector, thus under $\mathcal{M}_j$, only the $j$-th component of $\boldsymbol{\beta}$ is *not* flipped, and one has $\mathcal{M}_j \beta^i = -(-1)^{\delta_{ji}} \beta^i$. One can show that in systems with $\mathcal{M}_j$, the NLEE response $\chi_{ab}^{i,\beta}$ would vanish if $\delta_{ji} + \delta_{ja} + \delta_{jb}$ is an even number. Specific to monolayer MoTe$_2$, with in-plane electric field ($E_x$ or $E_y$), the only non-vanishing component of the NLEE tensor is $\chi_{xy}^{z,\beta}$, indicating that the magnetization induced by NLEE is along the out-of-plane direction. Note that if $\mathcal{M}_z$ is broken (e.g., by an electric field or in a Janus structure), then in-plane magnetization should exist.

Here we focus on CPL responses of monolayer MoTe$_2$ and plot $\xi_{xy}^{z,\beta}$ (Figure 2b). A prominent feature is that the orbital part $\xi_{xy}^{z,L}$ is about 25 times greater than the spin part $\xi_{xy}^{z,S}$. In other words, under CPL, the NLEE magnetization comes mostly from the orbital contribution, which is opposite to the behavior of equilibrium magnetization $M_0$ in magnetic materials, where $M_0^L \gg M_0^S$. This phenomenon can be better understood when we assume a sufficiently long relaxation time ($\hbar/\tau$ much smaller than the bandgap of the material) and use the two-band approximation, then we can simplify Eqs. (2,3) as [25]

$$\xi_{ab}^{i,\beta}(0;\omega,-\omega) \qquad (4)$$
$$= \tau \frac{\pi \mu_B e^2 V_{u.c.}}{2\hbar^2} \int \frac{d\boldsymbol{k}}{(2\pi)^3} \sum_{m \neq l} f_{lm}[r_{lm}^a, r_{ml}^b](\beta_{mm}^i - \beta_{ll}^i)\delta(\omega_{ml} - \omega)$$

Here $[r_{lm}^a, r_{ml}^b] = r_{lm}^a r_{ml}^b - r_{lm}^b r_{ml}^a$ is the interband Berry curvature, while $\Delta\beta_{ml}^i = \beta_{mm}^i - \beta_{ll}^i$ is the difference between the spin/orbital polarization on band $m$ and $l$. This formalism is illustrated on the right side of Figure 1b: light pumps transitions between bands $m$ and $l$, and the transition rate $R$ is determined by $R \propto [r_{lm}^a, r_{ml}^b]\delta(\omega_{ml} - \omega)$. The pumping process is compensated by the relaxation from band $l$ back to band $m$, which is characterized by the relaxation time $\tau$. In steady-state, the occupation number of conduction band $m$ is $\delta f_{m,l} \propto R\tau$. Eq. (4) simply states that the magnetization induced by light is $\delta M \propto \delta f_{m,l} \Delta\beta_{ml}^i$. MoTe$_2$ has a direct bandgap at K/K' points, and we schematically plot the band structure of MoTe$_2$ near the K valley in Figure 2c, while the K' valley can be similarly analyzed. The valance bands and conduction bands have major contributions from $d_{-2} = \frac{1}{\sqrt{2}}(d_{x^2-y^2} - id_{xy})$ and $d_0 = d_{z^2}$ orbitals of Mo atom, respectively [25]. Each of the valence and conduction bands is two-fold, and the degeneracy is broken by spin-orbit coupling (SOC). The Zeeman type spin splitting (up and down along $z$ direction) induced by SOC is indicated by the red (spin up) and blue (spin down) color. Note that the orbital character is mostly determined by crystal field, thus SOC does not significantly change it. There are four possible inter-band transitions, indicated by I-IV in Figure 2d. II and III have sizable $\Delta S^z$, while I and IV has $\Delta S^z \approx 0$. But $\Delta S^z$ of II is opposite to that of III, so the contributions from II and III tend to cancel each other and one has $\xi_{xy}^{z,S} \sim \text{II} - \text{III}$. As for the orbital part, all four transitions I-IV contributes to $\Delta L^z$, and their contributions are the same ($\Delta L^z \approx 2$ for $d_{-2} \to d_0$) and should be summed up, thus $\xi_{xy}^{z,L} \sim \text{I} + \text{II} + \text{III} + \text{IV}$. Furthermore, since II and III flip spin, their transition rate should be much lower than that of I and IV. Therefore, in general one would have $\xi_{xy}^{z,L} \gg \xi_{xy}^{z,S}$. In fact, light directly interacts with the orbital degree of freedom of the electrons, and leads to non-zero $\xi_{xy}^{z,L}$ with the orbital texture. The interaction is then transmitted to the spin degree of freedom by SOC, which leads to a finite $\xi_{xy}^{z,S}$ [27]. Also, the spin texture in MoTe$_2$ is created by SOC. Thus, one should naturally expect that the orbital contribution to the magnetization should be much greater if SOC is not too strong. Actually, without SOC, the spin-rotation symmetry is conserved, and the two valance bands and conduction bands are degenerate (no spin splitting). In this case, $\xi_{xy}^{z,S}$ vanishes, whereas $\xi_{xy}^{z,L}$ persists [25].

We now briefly compare the magnitudes of NLEE and LEE. The peak value of $\xi_{xy}^{z,T}$ is on the order of $10^3 \mu_B / \left(\frac{V}{\text{Å}}\right)^2$ (Figure 2c). We have also calculated the LEE response function $\zeta_a^i$ of MoTe$_2$, which exists only when the electron Fermi level $\mathcal{E}_F$ is tuned into the valence or conduction bands, so that the system becomes metallic. When $\mathcal{E}_F$ is 0.2 eV inside the valence or conduction bands, $\zeta_a^i$ is on the order of 0.1 $\sim$ 1 $\mu_B/\frac{V}{\text{Å}}$ [25]. This indicates that with $E \gtrsim 10$ MV/m, the NLEE strength would exceed that of LEE (indicated by $\xi E^2 > \zeta E$). Here we would like to remark that the NLEE magnetization can be detected by a pump-probe scheme: One first applies a strong pulsed laser to induce the magnetism in the system, then use a second weaker laser to detect the magnetism with magneto-optical effects, such as Faraday rotation, or Kerr rotation. Generally, the NLEE response tensor is on the order of $100\ \mu_B/\left(\frac{V}{\text{Å}}\right)^2$ (except for nonmagnetic materials under linearly polarized light). Therefore, when the electric field from the light is on the order of 1 V/m, the NLEE magnetization would be on the order of 1 $\mu_B$ per unit cell, which is magnitude-wise comparable with that of common magnetic materials, and is readily detectable. An electric field of 1 V/m corresponds to a light intensity of $1.3 \times 10^{11}$ W/cm$^2$, which is an experimentally accessible intensity, especially with pulsed lasers.

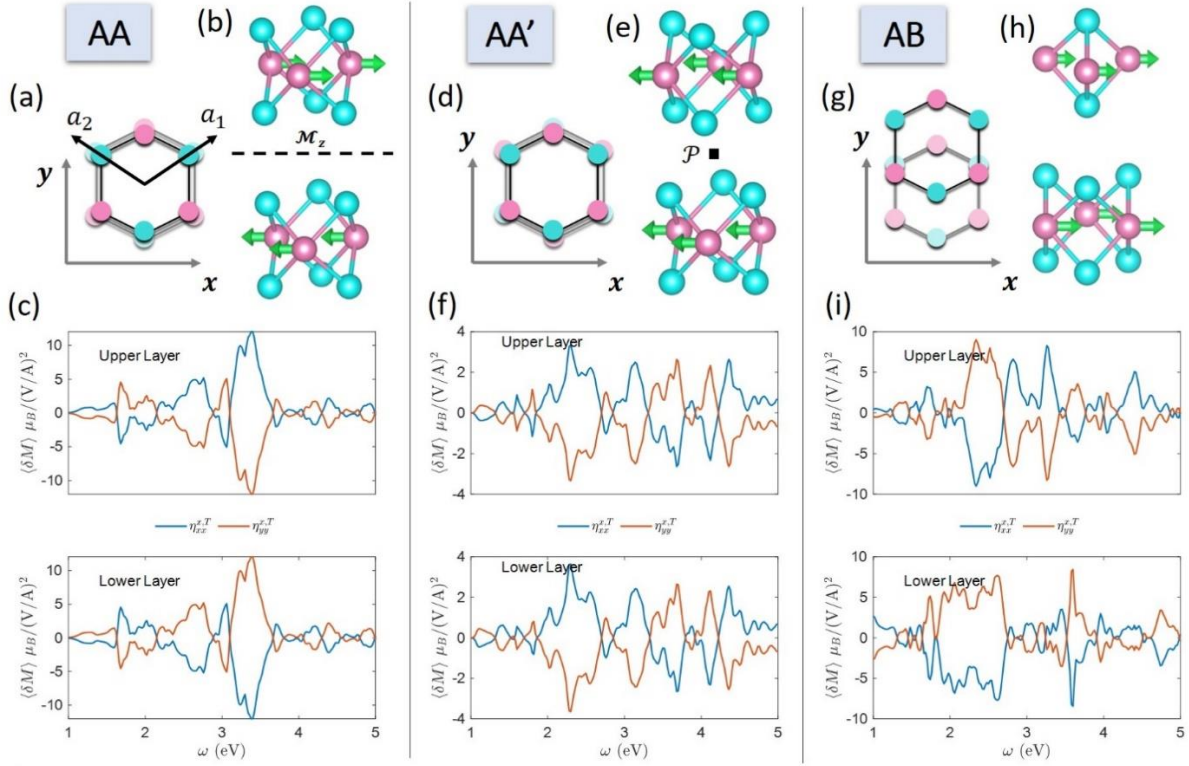

**Figure 3** The nonlinear Edelstein effect under linearly polarized light of bilayer MoTe$_2$ with AA (left), $AA'$ (middle) and AB (right column) stacking patterns. (a,d,e) are skematic plots of the stacking pattern. (b,e,h) show the magnetic order under linearly polarized light with the green arrows indicate the magnetization from the nonlinear Edelstein effect. Pink: Mo; Cyan: Te. (c,f,i) show the response function $\eta_{xx}^{x,T} = -\eta_{yy}^{x,T}$ for the upper layer and the low layer.

**Bilayer MoTe$_2$: Stacking dependent magnetic orders.** As described in the previous section, the NLEE magnetization of monolayer MoTe$_2$ is the same for all unit cells, which can be considered as in-plane FM ordering. On the other hand, in multi-layer or multi-sublattice systems, the different layers or sublattices may have different chemical/structural environments, and the local NLEE magnetizations associated with these layers (sublattices) do not have to be the same. As a result, various magnetization orderings, including AFM and FM, can be realized.

Here we use bilayer MoTe$_2$ as an example. Two monolayer MoTe$_2$ are stacked along the $z$ direction, and there can be many different stacking patterns. Three high symmetry stacking patterns of bilayer MoTe$_2$ are shown in Figure 3. In AA stacking (Figure 3a), Mo (Te) atoms of the upper layer sit directly above Mo (Te) atoms of the lower layer, and the two layers are mirror images of each other, with a horizontal mirror plane $\mathcal{M}_z$ (dash line in Figure 3a). In AA′ stacking (Figure 3d), Mo (Te) atoms in the upper layer are above the Te (Mo) atoms in the lower layer, and there is an inter-layer inversion symmetry $\mathcal{P}$, and the inversion center is indicated by the black box in Figure 3d. Finally, the AB stacking (Figure 3g) can be obtained by shifting the upper layer of the AA stacking by a vector of $\frac{1}{3}(a_1 + a_2)$, where $a_1$ and $a_2$ are lattice vectors (Figure 3a). Note that AB stacking has neither $\mathcal{M}_z$ nor $\mathcal{P}$. According to our first-principles calculations, the AA′ configuration is the most stable with the lowest energy, and AB has slightly higher energy (0.018 eV per unit cell), whereas AA has much higher energy (0.163 eV per unit cell).

Although the van der Waals interaction between the two MoTe$_2$ layers is weak, the stacking pattern strongly affects NLEE magnetization pattern. Here we calculate the layer-resolved (see Methods) response function $\eta_{xx}^{x,T} = -\eta_{yy}^{x,T}$ for all three stacking patterns. In Figure 3, $\eta$ for the upper (lower) layer corresponds to the NLEE magnetization of the upper (lower) layer under light. One can see that the NLEE magnetization patterns are distinct for the three stacking patterns: For AA stacking, $\eta^T$ on the upper and

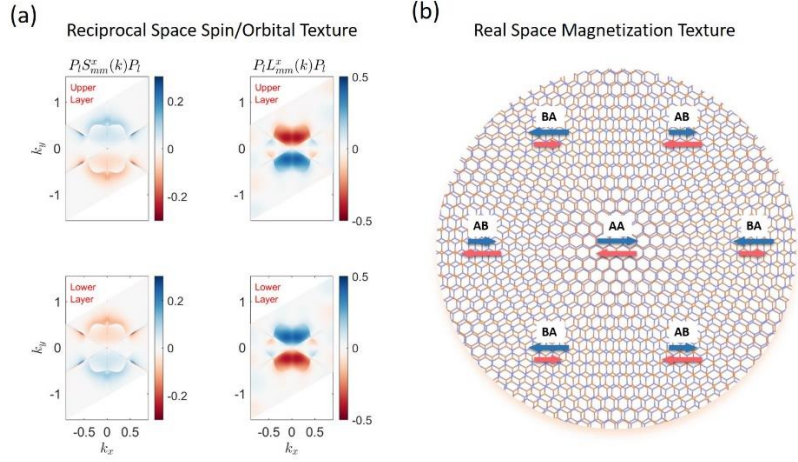

**Figure 4** (a) Reciprocal space local spin/orbital texture $P_l \beta^x_{mm}(\boldsymbol{k}) P_l$ for the highest valance band on the upper and lower layer of bilayer MoTe$_2$. The left (right) column is the spin (orbital) texture. (b) Real space magnetzation texture on a Moire pattern. The blue and pink arrows indicate magnetization on the upper and lower layers, respectively.

lower layers are exactly opposite (Figure 3c), thus under light illumination, the NLEE magnetization on the upper and lower layers would be anti-parallel, which is AFM ordering. Note that the total magnetization of the upper and lower layers is exactly zero, but the local magnetization on each layer does exist, reminiscent of the AFM order. For AA′ stacking, the two layers exhibit parallel magnetization ($\eta^T$ on the upper and lower layers are equal, Figure 3f), which can be considered as an FM ordering. Finally, for the AB stacking, $\eta^T$ on the upper and lower layers do not exhibit a simple relationship (Figure 3i), and different NLEE magnetization on upper and lower layers (staggered magnetism) is thus expected. The magnetic orderings of different stacking patterns come from symmetry constraints. For example, in AA stacking the inter-layer mirror operation $\mathcal{M}_z$ swaps the two layers and flips $M^x$. Consequently, local $M^x$ associated with the two layers must be the opposite to preserve the mirror symmetry of AA stacking. Actually, this effect is also manifested in the layer-projected $\boldsymbol{k}$-space spin/orbital texture $P_l \beta^x_{mm}(\boldsymbol{k}) P_l$ (see Methods), as shown in Figure 4a for the highest valence band. One can see that for any $\boldsymbol{k}$-point, the textures on the upper and lower layers are exactly opposite to each other. In equilibrium states, the spin/orbital polarization of all occupied states sum up to zero, hence local spin/orbital polarizations are *hidden* [28]. However, when the system is driven out of equilibrium, the hidden magnetization would emerge, and an AFM magnetization appears. Similar reasonings apply to AA′ stacking pattern, where the inversion symmetry $\mathcal{P}$ enforces FM ordering. As for AB stacking, there are no inter-layer symmetry constraints, hence the magnetizations on the two layers are not directly correlated. Interestingly, when the two layers are twisted to form a Moiré pattern, a real-space spin texture can be created. The Moiré pattern has spatially varying stacking patterns, which leads to spatially varying magnetic orderings with NLEE (Figure 4b). Besides, the non-equilibrium magnetization can be either (anti-)parallel or perpendicular to the electric field, which may lead to interesting physical phenomena. Particularly, the (anti-)parallel electric and magnetic field can be regarded as a nonlinear axion coupling [29].

**CrI$_3$: AFM Order Manipulation.** Until now, we have been discussing non-magnetic materials, where the spontaneous magnetization $M_0$ is zero, and a non-equilibrium magnetization $\delta M$ is generated under light. This can be considered as a non-magnetic to magnetic transition. On the other hand, in magnetic materials, there is already finite spontaneous magnetization $M_0$ in equilibrium. Light illumination could induce an additional NLEE magnetization $\delta M$. This $\delta M$ can be considered as an effective magnetic field $H_{\text{eff}}$, which exerts torques on $M_0$. Previous studies based on LEE suggest that this $H_{\text{eff}}$ can cause the precession of magnetic moments, and a magnetic phase transition may occur when $H_{\text{eff}}$ is strong enough [4,30,31]. Recently, the AFM spintronics [7,32,33] has attracted great interest. Compared with FM materials, using AFM materials has several advantages, such as the insensitivity to external magnetic fields, the absence of stray fields, and the fast dynamics with terahertz frequency, etc. Manipulating the magnetic ordering of AFM materials requires that the torque on the two magnetic sublattices are opposite so that no net magnetization is induced. Obviously, this cannot be achieved with a static external magnetic field. A few approaches have been proposed to manipulate AFM ordering, such as electrical approaches based on LEE [5,12,34], and optical approaches [7] based on IFE [35].

Here we propose that NLEE can be an alternative methodology for manipulating AFM ordering. Compared with LEE, NLEE applies to semiconductors, and the choice of light frequency, polarization, and intensity could provide good flexibility. Furthermore, the ultra-fast ultra-strong pulsed lasers render it possible to manipulate the AFM ordering, and even trigger AFM order switching (i.e., $M_0 \to -M_0$) on a picosecond timescale. To illustrate the NLEE in magnetic materials, we take bilayer CrI$_3$ as an example. The magnetic ground state of bilayer CrI$_3$ is AFM with the magnetization $M_0$ along the $z$ direction [36]. To be specific, we assume that $M_0$ on the upper (lower) layer point downwards (upwards, inset of Figure 5a). The layer-resolved response functions $\eta$ and $\xi$ under this configuration are plotted in Figure 5, where a positive (negative) value of $\eta/\xi$ indicates a $\delta M = \eta E^2/\xi E^2$ along $+z$ ($-z$) direction. One can see the under CPL, $\delta M$ on the two layers are parallel to each other, whereas under LPL they are (approximately) anti-parallel. Notably, for $\omega \gtrsim 3$ eV, $\eta$ on the upper (lower) layer is negative (positive), so $\delta M$ is opposite to $M_0$, which can be utilized to swap the magnetization and trigger an AFM order switching. The opposite NLEE magnetization on the upper and lower layers are beneficial since it would keep the total magnetization zero, so that the system stays AFM. We estimate the effective magnetic field from $H_{\text{eff}} = \frac{\delta M J_{\text{ex}}}{(\mu_{\text{B}})^2}$, where $J_{\text{ex}}$ is the exchange energy between carrier spin and the local magnetic moment, and is estimated to be ~1 eV from band structures (see [25]: for an order-of-magnitude estimation, here we assume that carrier orbital and spin magnetization have identical exchange energy with local magnetic moment). From Figure 5a one can see that $\eta$ is on the order of $100\ \mu_{\text{B}}/\left(\frac{V}{Å}\right)^2$, yielding $H_{\text{eff}} \sim 10^6\ \text{T}/\left(\frac{V}{Å}\right)^2$. Therefore, an electric field $E \sim 0.1$ MV/cm (corresponding to light intensity $\sim 27$ MW/cm$^2$) could generate an $H_{\text{eff}} \sim 1$ T, which is strong enough coercive field to trigger a magnetic-order transition in CrI$_3$ [3]. The temperature increase under such light illumination is estimated to be on the order of 10 K [25], thus CrI$_3$ can be kept below its Néel temperature, which is around 45 K [36,37].

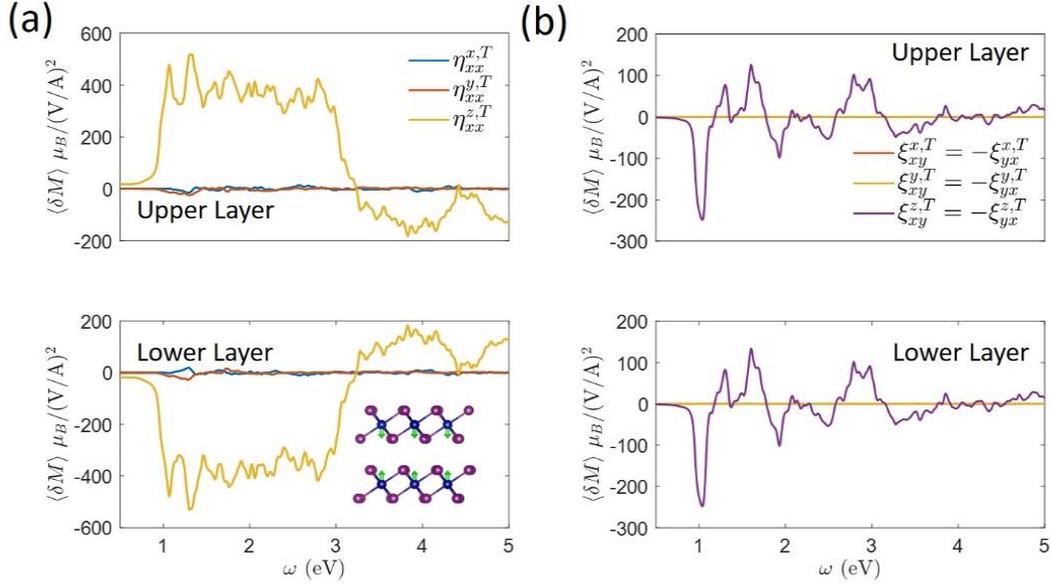

**Figure 5** Nonlinear Edelstein effect of anti-ferromagnetic CrI$_3$ with magnetization along $z$ axis. Under (a) linearly and (b) circularly polarized light, the light induced magnetization are opposite and parallel on the two layers, respectively. Inset of (a): atomic strcuture of bilayer CrI$_3$. The green arrows indicate the equilibrium magnetization.

## Discussion

First, we would like to discuss the relationship between NLEE and IFE [38–40] and ICME [41–43], which also generate an effective magnetic field $H^{\text{eff}}$ under CPL and LPL, respectively. The light-matter interaction through the electric field can be described by the Hamiltonian $\mathcal{H}_{\text{int}} = \sum_{ab} \frac{1}{2} \varepsilon_{ab} E_a E_b^*$, where $\varepsilon_{ab}$ is the dielectric function. Phenomenologically, IFE and ICME come from the derivative of $\mathcal{H}_{\text{int}}$ with respect to magnetization, i.e., $H_k^{\text{eff}} = \frac{\partial \mathcal{H}_{\text{int}}}{\partial M_k} = \sum_{ab} \frac{1}{2} \frac{\partial \varepsilon_{ab}}{\partial M_k} E_a E_b^*$. The dielectric function $\varepsilon_{ab}$ depends on the magnetic state of the system. Due to the symmetry constraints [44], to the lowest order $\varepsilon$ satisfies $\varepsilon_{ab}^{(a)} = \sum_k \alpha_{abk} M_0^k$ and $\varepsilon_{ab}^{(s)} = \varepsilon_{ab}^0 + \frac{1}{2} \sum_{kl} \beta_{abkl} M_0^k M_0^l$, where $\varepsilon^0$ is vacuum permittivity, while $\varepsilon^{(a)}/\varepsilon^{(s)}$ are the asymmetric/symmetric part of $\varepsilon$. $\alpha_{abk}$ and $\beta_{abkl}$ are phenomenological parameters. Thus, after the derivative with respective to $M$ one has $H_k^{\text{IFE}} \propto \sum_{ab} \alpha_{abk}(E_a E_b^* - E_b E_a^*)$ and $H_k^{\text{ICME}} \propto \sum_{abl} \beta_{abkl} M_0^l (E_a E_b^* + E_a E_b^*)$. Consequently, in non-magnetic materials ($M_0 = 0$), IFE can exist, while ICME must vanish. On the other hand, NLEE, which generates non-equilibrium magnetization $\delta M$, can be understood as the derivative of $\mathcal{H}_{\text{int}}$ with respect to magnetic field, i.e., $\delta M^k = \sum_{ab} \frac{1}{2} \frac{\partial \varepsilon_{ab}}{\partial H_k} E_a E_b^*$. Since $H$ and $M$ are conjugate variables, NLEE and IFE/ICME can be regarded as two complementary perspectives on the same magneto-optic effect. Notably, our quantum theory provides an approach to calculate the response function $\chi_{ab}^{i,\beta}$ with *ab initio* calculation, whereas $\alpha_{abk}$ and $\beta_{abkl}$, to the best of our knowledge, cannot yet be calculated directly. In addition, we clarify that the magnetization $\delta M$, or equivalently the effective field $H^{\text{eff}}$, can be generated under LPL in nonmagnetic materials if the frequency of the light is above the bandgap of the material, in contrast to the conclusion from the phenomenological analysis above, which suggests that $H_k^{\text{ICME}}$ should be zero when $M_0^l$ is zero. The reciprocity is broken by energy dissipation and the NLEE can be regarded as a non-reciprocal process.

Second, the spin dynamics of AFM materials such as CrI3 under NLEE remains to be studied. Note that under LPL, one (approximately) has $\delta M^k \propto H_k^{\text{ICME}} \propto \sum_{abl} \beta_{abkl} M_0^l (E_a E_b^* + E_a E_b^*)$. When the magnetic anisotropy is not too strong, it is reasonable to assume that the off-diagonal terms ($k \neq l$) of $\beta_{abkl}$ are much smaller the diagonal terms ($k = l$), thus $H$ should be approximately (anti-)parallel to $M_0$, which is verified by our *ab initio* calculations [25]. The spin dynamics under $H$ with such a pattern shall be studied carefully to determine whether it is possible to trigger AFM order switching, and if possible, to determine the optimal light pulse intensity, polarization and duration.

**Conclusion**

In conclusion, we have developed a quantum theory of the nonlinear Edelstein effect, which is the generation of magnetization under light illumination. Based on symmetry analysis, we demonstrate that the NLEE is not constrained by either spatial inversion symmetry or time-reversal symmetry, and is thus widely applicable to many materials systems in a non-contact manner. Particularly, we elucidate that orbital- and spin-magnetization could emerge even in conventionally nonmagnetic materials under linearly polarized light, which is counter-intuitive. We attribute this to the breaking of time-reversal symmetry by the energy dissipation of photocurrents under light illumination. Then we demonstrate various features of the NLEE. First, we illustrate that the contribution from the orbital degree of freedom to the total NLEE magnetization can be much higher than that from the spin degree of freedom, which is opposite to the common notion for equilibrium (intrinsic) magnetizations. Then using bilayer MoTe2 as an example, we show that different opto-magnetic orderings, including ferromagnetic, anti-ferromagnetic orderings, are realizable in multi-layer or multi-sublattice systems, depending on the symmetries that the system possesses. Finally, with bilayer CrI3 as an example, we demonstrate that the magnetization induced by NLEE may also effectively manipulate magnetic ordering in semiconducting and insulating magnetic materials, unlike the linear Edelstein effect which is applicable only in metals. Magnitude-wise, the NLEE can lead to larger magnetization than the LEE when the electric field strength is greater than 10 MV/m. Experimentally, a (pulsed) laser with electric field strength on the order of 1 V/nm would be able to generate a magnetization on the order of 1 $\mu_\text{B}$ per unit cell, which is readily detectable. The NLEE provides a convenient way to generate magnetization with light, and may find applications in e.g., ultrafast spintronics and quantum information processing.

## Methods

### Density functional theory and Wannier Calculations

The Vienna *ab initio* simulation package (VASP) [45,46] is used for the first-principles calculations based on density functional theory (DFT) [47,48]. The exchange-correlation interactions are treated by the generalized gradient approximation (GGA) in the form of Perdew-Burke-Ernzerhof (PBE) [49]. Projector augmented wave (PAW) method [50] and plane-wave basis functions are used to treat the core and valence electrons, respectively. For DFT calculations, the first Brillouin zone is sampled by a Γ-centered ***k***-mesh with a grid of $25 \times 25 \times 1$ for MoTe$_2$ and $15 \times 15 \times 1$ for CrI$_3$. The DFT+$U$ method is adopted to treat the $d$ orbitals of spin polarized Cr atoms in CrI$_3$ ($U = 3.0$ eV). Tight-binding orbitals are generated from Bloch waves in DFT calculations, using the Wannier90 package [51]. The tight-binding Hamiltonian is then used to interpolate the band structure on a much denser ***k***-mesh calculate the LEE and NLEE response functions, which is $512 \times 512 \times 1$ for MoTe$_2$, and $320 \times 320 \times 1$ for CrI$_3$. The $\mathbf{k}$-mesh convergence is well tested. In order to calculate the real space local magnetization, we define a projection operator $P_l = \sum_{i \in l} |\psi_i\rangle\langle\psi_i|$, where $l$ denotes the spatial region (e.g., $l$-th layer in a multi-layer system, or the $l$-th sublattice in a multi-sublattice system), while $|\psi_i\rangle$ is the tight binding orbital belonging to region $l$. Then the operator $P_l \beta P_l$ is to replace $\beta$ operator in Eq. (2). We simply used atomic orbitals (s, p, d, etc.) to calculate the orbital angular momentum ($\beta = L$). So only the contribution from intra-atom term $\langle n\mathbf{R}|\mathbf{r} \times \mathbf{p}|m\mathbf{R}\rangle$ is included, while the contribution from inter-atom term $\langle n\mathbf{R}|\mathbf{r} \times \mathbf{p}|m\mathbf{R}'\rangle$ is neglected. The model assumptions of (a) a uniform carrier lifetime $\tau$ for all electronic states, and (b) only atomic orbitals (s, p, d, etc.) $\langle n\mathbf{R}|\mathbf{r} \times \mathbf{p}|m\mathbf{R}\rangle$ contribute to the total orbital angular momentum, while neglecting inter-atomic $\langle n\mathbf{R}|\mathbf{r} \times \mathbf{p}|m\mathbf{R}'\rangle$ contributions, can certainly be systematically improved in future works.

**Data availability**

The MATLAB code for calculating the NLEE magnetization is available at http://alum.mit.edu/www/liju99/NLEE

**Acknowledgments**

This work was supported by an Office of Naval Research MURI through grant #N00014-17-1-2661. We are grateful for discussions with Zhurun Ji and Hanyu Zhu.

**Conflict of interests**

The authors declare no competing financial or non-financial interests.

**Contributions**

J.L. and H.X. conceived the idea and designed the project. H.X. performed the calculations with the help of J.Z. and H.W. J.L. supervised the project. All authors analyzed the data, wrote the paper, and contributed to the discussions of the results.